\documentclass[reprint, superscriptaddress]{revtex4-1}
\usepackage[T1]{fontenc}
\usepackage[english]{babel}
\usepackage{amsmath,amsfonts,amsthm}
\usepackage{graphicx}
\usepackage{xcolor}
\usepackage{pgfplots}
\usepackage{anyfontsize}
\usepackage{calc}
\usepackage{tikz}
\usepackage{braket}
\usepackage{lipsum}
\usepackage{wrapfig}
\usepackage[margin=0.75in]{geometry}
\usepackage{pdfpages}
\usepackage{fancyhdr}

\makeatletter
\AtBeginDocument{\let\LS@rot\@undefined}
\makeatother
\pgfplotsset{compat=newest}
\usetikzlibrary{arrows}
\usetikzlibrary{calc}

\renewcommand{\vec}[1]{\mathbf{#1}}

\numberwithin{equation}{section} 
\numberwithin{figure}{section} 
\numberwithin{table}{section}

\DeclareMathOperator{\Span}{span}

\begin{document}

\title{ Error Mitigation in Quantum Computers subject to Spatially Correlated Noise  }
\author{Vickram N. Premakumar}
\address{Physics Department, University of Wisconsin-Madison, 1150 Univ. Ave., Madison, WI, USA}
\author{Robert Joynt}
\address{Physics Department, University of Wisconsin-Madison, 1150 Univ. Ave., Madison, WI, USA}
\address{
Kavli Institute for Theoretical Sciences, University of Chinese Academy of Sciences, Beijing 100190, China}
\date{{\normalsize \today}}

\begin{abstract}
The most common error models for quantum computers assume the independence
of errors on different qubits. However, most noise mechanisms have some
correlations in space. We show how to improve quantum information
processing for few-qubit systems when spatial correlations are present. This
starts with strategies to measure the correlations. Once the correlations
have been determined, we can give criteria to assess the suitability of
candidate quantum circuits to carry out a given task. \ This is achieved by
defining measures of decoherence that are local in Hilbert space,
identifying ``good'' and ``bad'' regions of the space. \ Quantum circuits that
stay in the ``good'' regions are superior. \ Finally, we give a procedure by 
means of which the improvement of few-qubit systems can be extended to large-scale
quantum computation. \ The methods described here work best when dephasing noise dominates
over other types of noise. \ The basic conceptual theme of the work is the
generalization of the concept of decoherence-free subspaces in order to
treat the case of arbitrary spatial correlations.
\end{abstract}

\maketitle


\section{Introduction}

The most basic justification of the pursuit of quantum computation is the
existence of threshold theorems. They tell us that if a certain precision at
the qubit level can be achieved, then a workable quantum computer (in
principle of arbitrary size) can be made. These thresholds also give
concrete goals for hardware performance in systems containing only a few
qubits. However, threshold theorems generally make the key assumption that
errors on different qubits occur in statistically independent fashion \cite%
{Preskill2013}. Some relaxation of this condition can be allowed \cite%
{Preskill1998, Bombin2016}, but error correction then becomes more complicated and
resource-intensive. On the other hand, it is known that correlations in
noise can actually be used to fight noise-induced degradation in
performance, using the concept of decoherence-free subspaces (DFS) \cite%
{Lidar1998, Duan1998, Lidar2001, Lidar2001_2}. The resources involved in utilizing DFS
appear to be less than in most error-correction schemes.

This situation raises some interesting questions. If the noise is
correlated, is this fundamentally good or bad for quantum computation? If we
have the choice of dealing with the situation by error correction or DFS,
which is less expensive? If we do not have a DFS, but there are some
correlations in the noise, is it still possible to reduce the computation's susceptibility to noise by appropriate protocols?

We investigate these questions here in two stages. \ We begin by looking at
a small system of only two physical qubits.  With very small systems such as
this, quantum error correction is of course out of the question. \
Furthermore, the use of a DFS would prohibit any nontrivial quantum
information processing. \ We will show, however, that significant error%
\textit{\ mitigation} is still possible.  In an era when quantum computing resources are not nearly sufficient for true error correction, this is an appropriate subject of research \cite{Temme2017}. \ We will describe methods that may
be used for any such system. For illustration purposes, we will use a
concrete experimental example: two electron spin qubits in a Si/SiGe
heterostructure \cite{Watson2018}. This system has the great advantage that
it is completely programmable, and we have some insight into the types of
noise to be expected \cite{Friesen2017, Zwanenburg2013}. \ In the second
stage of our work, we extend the concepts used for two-qubit systems to
many-qubit systems, as far as possible. \ The overall aim is to understand
how to improve quantum information processing when noise correlations are
present.

Sec. II introduces the model and constructs the framework to describe the
spatial and time correlations in the noise, limiting the discussion to
dephasing noise. \ In Sec. III we propose a measurement scheme to obtain
these correlations, which is a simple and easily understood extension of
methods used for single qubits \cite{Biercuk2009, Alvarez2011,  Yuge2011}. It can also be viewed as a concrete
application of much more general schemes given in some recent papers \cite{Alvarez2011, Sza2016, Paz-Silva2017, Krzywda2018}.\ We apply these methods to the specific
system in question to obtain the important auto- and cross-correlation
functions. \ In Sec. IV, we generalize DFS concepts to obtain measures of
decoherence that are local in Hilbert space and use these measures to show
how to improve the robustness of a quantum circuit. \ This increases the
fidelity obtained for a given quantum information processing task. \ In Sec. V we apply the method to the model two-qubit system and discuss the advantages and limitations of our recommendations for error mitigation. \ Sec. VI treats the extensions to many-qubit systems and and addresses the scalability of the method. \   In Sec. VII we conclude by considering possible generalizations to other error models and how to combine our method with quantum error correction.

\section{Noise Correlations}

The model Hamiltonian for two qubits subject to dephasing noise is 
\begin{equation*}
H=H_{0}+H_{g}\left( t\right) +H_{n}\left( t\right) .
\end{equation*}%
It consists of 
\begin{equation*}
H_{0}=b_{1}Z_{1}+b_{2}Z_{2},
\end{equation*}%
a static Hamiltonian that provides the qubit splittings $b_{1}$ and $b_{2}$
in energy units. ($b_{1}$ and $b_{2}$ need not be magnetic fields.) $%
H_{g}\left( t\right) $ is the gate Hamiltonian that is used to do qubit
operations. \ $X_{i},Y_{i},Z_{i}$ are the Pauli matrices on site i. The noise Hamiltonian is
\begin{equation*}
H_{n}\left( t\right) =\delta b_{1}\left( t\right) Z_{1}+\delta b_{2}\left(
t\right) Z_{2}.
\end{equation*}%
In the case of electron spins in an inhomogeneous
magnetic field, this choice of $H$ models random electric fields that move
the qubits and vary their splittings. \ 
We use the product basis $\left\{ \left\vert 00\right\rangle ,\left\vert
01\right\rangle ,\left\vert 10\right\rangle ,\left\vert 11\right\rangle
\right\} $, which also forms an eigenbasis for $H_{0}$. 
The single-qubit dephasing times $T_{2}^{\left( 1\right) }$ and $%
T_{2}^{\left( 2\right) }$ for qubits 1 and 2 are determined by the local
noise spectra 
\begin{equation*}
S_{ij}\left( \omega \right) =\int_{-\infty }^{\infty }\left\langle \delta
b_{i}\left( t\right) ~\delta b_{j}\left( 0\right) \right\rangle \cos \omega
t~dt\equiv \left\langle \delta b_{i}~\delta b_{j}\right\rangle _{\omega }
\end{equation*}%
on the two qubits. \ In the simplest theory \cite{Slichter2010} we have 
\begin{equation*}
\frac{1}{T_{2}^{\left( j\right) }}=\frac{4}{\hbar ^{2}}\lim_{\omega
\rightarrow 0}S_{jj}\left( \omega \right) 
\end{equation*}%
as long as the longitudinal relaxation time $T_{1}^{\left( j\right)
}>>T_{2}^{\left( j\right) },$ which is usually the case. \ More accurate
formulas can be used, but they do not change the basic physics that $T_{2}$
comes from the noise spectrum at low frequencies. 

We propose an experiment to measure \ 
\begin{equation*}
S_{12}\left( \omega \right) =\int_{-\infty }^{\infty }\left\langle \delta
b_{1}\left( t\right) ~\delta b_{2}\left( 0\right) \right\rangle \cos \omega
t~dt\equiv \left\langle \delta b_{1}~\delta b_{2}\right\rangle _{\omega }.
\end{equation*}%
This is a correlation function of the noise at different spatial locations.
\ 

Measuring $S_{12}\left( \omega \right) $ is of interest for two reasons.

First, it tells us something about the nature of the noise. \ For example,
in semiconductor implementations with charge qubits, charge noise is often
the dominant decoherence mechanism \cite{Yoneda2018}. \ If this is due to defects that are far
from the qubits, then the noise from the random electric field has
wavelengths much longer than the separation of the qubits, and the random
electric field is about the same at the two qubits. The opposite limit is when the
defect lies between the qubits when we expect anticorrelation in the
electric field. \  \ 

Second, we can use the information to design noise-resistant operations,
which is the focus of this paper. \ If $\left\langle \delta b_{1}~\delta
b_{2}\right\rangle_\omega $ is large and positive, then $\left\langle \left( \delta
b_{1}+\delta b_{2}\right) \left( \delta b_{1}+\delta b_{2}\right)
\right\rangle_\omega $ is large and $\left\langle \left( \delta b_{1}-\delta
b_{2}\right) \left( \delta b_{1}-\delta b_{2}\right) \right\rangle_\omega $ is
small. \ Looking back at the noise Hamiltonian
for this model we see 
\begin{equation*}
H_{n}\left( t\right) \approx \delta b_{1}\left( t\right) \left(
Z_{1}+Z_{2}\right),
\end{equation*}%
which only couples to $Z_{tot}=Z_{1}+Z_{2}.$ \ This means that the subspace
spanned by $\left\{ \left\vert 01\right\rangle ,\left\vert 10\right\rangle
\right\} $ is approximately a decoherence-free subspace. \ Conversely, if If 
$\left\langle \delta b_{1}~\delta b_{2}\right\rangle_\omega $ is large and
negative, then 
\begin{equation*}
H_{n}\left( t\right) \approx \delta b_{1}\left( t\right) \left(
Z_{1}-Z_{2}\right) 
\end{equation*}%
and the subspace spanned by $\left\{ \left\vert 00\right\rangle ,\left\vert
11\right\rangle \right\} $ is approximately a decoherence-free subspace. \
By working ``near'' the appropriate subspace we can get a lower error rate. \
This is essentially the same idea as singlet-triplet qubits \cite{Petta2005}, where only two
levels are used to define a single logical qubit. \ Our aim here is quite
different. \ We keep the full 2-qubit system and see if we can use the noise
correlations to help design a small quantum information processing device. \
Since the ultimate aim of the paper is to improve the performance of the device, we limit our focus in what follows to correlations that are most likely to lead to simple usable DFSs. \ As we will see,
this limitation also means that out methods work well only when dephasing noise dominates over other noise.

Clearly, the definition of $S_{ij}\left( \omega \right) $ generalizes
immediately to multiple qubits. \ We can define cross-correlation functions
for any pair and it may well happen that only short-range pairwise
correlations are important. \ The usefulness of such generalizations will be
discussed further in Sec. VI.

\section{Measuring Correlations}

In this section we suggest an experimental strategy to determine the spatial
correlations. \ They are determined by means of a measurement analogous to
the measurement of Ramsey fringes. In the 4-dimensional two-qubit space we
may choose any 2-dimensional subspace to perform the measurement.  However, as we have noted, some subspaces are more likely than others to be DFSs, and these are the most likely to be of real usefulness.
Hence we will focus on the two subspaces that are DFSs when the noise is perfectly correlated and when it is perfectly anticorrelated.

\subsection{\textbf{Experiment 1. \ Ramsey in the }$\left\{ \left\vert
00\right\rangle ,\left\vert 11\right\rangle \right\} $\textbf{\ basis.}}

Let the north pole of a Bloch sphere be$~\left\vert 00\right\rangle $ and
the south pole be $\left\vert 11\right\rangle .$ \ We start in the state $%
\left\vert 00\right\rangle $ and then use $H_{g}$ to make a $\pi /2$
rotation about the y-axis preparing the state%
\begin{equation*}
\Psi _{+}\left( t=0\right) =\frac{1}{\sqrt{2}}\left( \left\vert
00\right\rangle +\left\vert 11\right\rangle \right) ,
\end{equation*}%
and then let it evolve under the influence of $H_{0}$ alone$.$ \ Then we
have 
\begin{equation*}
\Psi _{+}\left( t\right) =\frac{1}{\sqrt{2}}e^{-i\left( b_{1}+b_{2}\right)
t}\left\vert 00\right\rangle +\frac{1}{\sqrt{2}}e^{i\left(
b_{1}+b_{2}\right) t}\left\vert 11\right\rangle ,
\end{equation*}%
and if we make another $\pi /2$ rotation about the y-axis and then measure
the probability of being in the state $\left\vert 11\right\rangle $ after a
time $t$ we get 
\begin{equation*}
P_{+} 
=\frac{1}{2}+\frac{1%
}{2}\cos \left[ 2\left( b_{1}+b_{2}\right) t\right] ,
\end{equation*}%
so the period is $\tau _{+}=\pi /\left( b_{1}+b_{2}\right) .$ \ The
experiment is illustrated in Fig. 1. \ If we now add in $H_{n},$ the noise,
we find 
\begin{equation*}
P_{+}=\frac{1}{2}+\frac{1}{2}e^{-t/T_{2}^{\left( +\right) }}\cos \left[
2\left( b_{1}+b_{2}\right) t\right] ,
\end{equation*}%
in a certain time window longer than the inverse cut-off time of the noise.
\ (At shorter times the decay is Gaussian.) \ Here $1/T_{2}^{\left( +\right)
}$ is given by the integral of the Fourier transform of $4\left\langle \left[
 \left( \delta b_{1}+ \delta b_{2}\right) \right] ^{2}\right\rangle_\omega $, a windowing
function that depends on the approximation being used, and some other
factors involving the temperature, $\hbar ,$ etc. \ Omitting these
prefactors and others that depend on the precise form of the power spectrum
we have that 
\begin{equation*}
1/T_{2}^{\left( +\right) }\sim \left\langle \left[ \left(
\delta b_{1}+ \delta b_{2}\right) \right] ^{2}\right\rangle_\omega .
\end{equation*}%
The number of oscillations observed will be $N_{+}$, which is%
\begin{equation*}
N_{+}=\frac{T_{2}^{\left( +\right) }}{\tau _{+}}\sim \frac{b_{1}+b_{2}}{4\pi
\left\langle \left[  \left( \delta b_{1}+ \delta b_{2}\right) \right]
^{2}\right\rangle_\omega }.
\end{equation*}

In the case of perfectly anticorrelated collective dephasing $N_{+}$ diverges, a signature of
a perfect DFS.

\begin{figure}[t]
\includegraphics[width=1\linewidth]{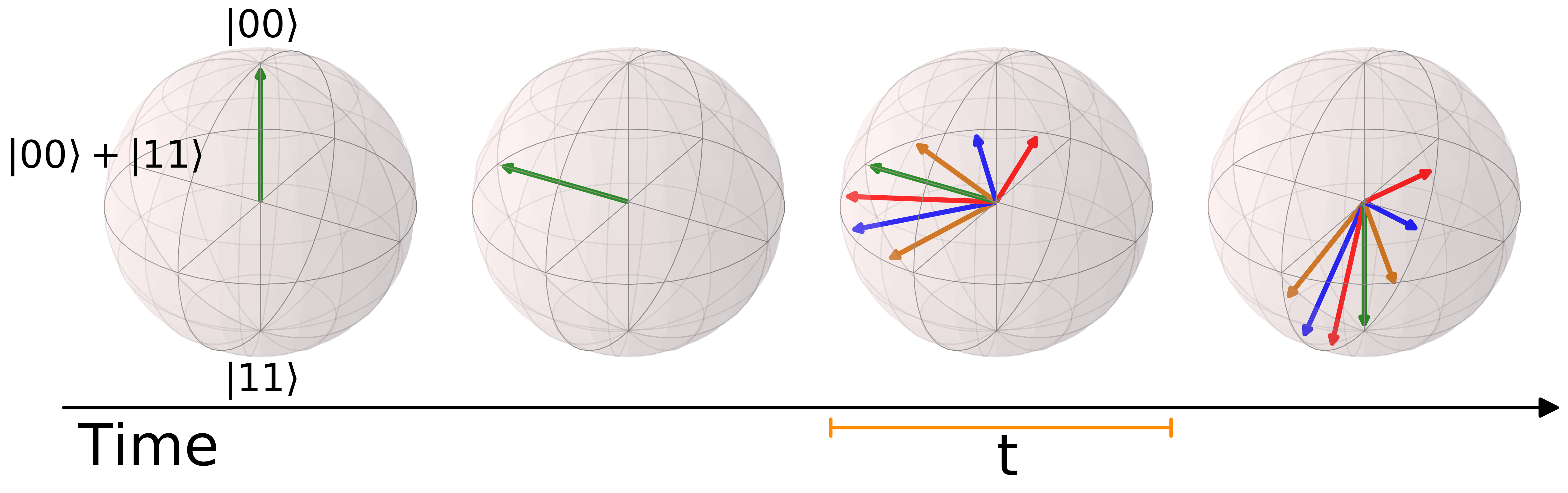}
\caption{Illustration of the proposed Ramsey-type experiments to measure
noise correlations in a Bloch-sphere representation. $T_{2}^{\left( +\right)
}$ and $T_{2}^{\left( -\right) }$ are determined by the time decay of the
amplitude on the South Pole in Experiments 1 and 2 respectively.}
\label{fig:ramseybloch}
\end{figure}

\subsection{\textbf{Experiment 2. \ Ramsey in the }$\left\{ \left\vert
01\right\rangle ,\left\vert 10\right\rangle \right\} $ \textbf{basis.}}

We start in the state $\left\vert 01\right\rangle $ and then use $H_{g}$ to
make a $\pi /2$ rotation about the y-axis in the $\left\{ \left\vert
01\right\rangle ,\left\vert 10\right\rangle \right\} $ subspace preparing
the state%
\begin{equation*}
\Psi _{-}\left( t=0\right) =\frac{1}{\sqrt{2}}\left( \left\vert
01\right\rangle +\left\vert 10\right\rangle \right) .
\end{equation*}%
We have 
\begin{equation*}
\Psi _{-}\left( t\right) =\frac{1}{\sqrt{2}}e^{ -i\left( b_{1}-b_{2}\right)
t} \left\vert 01\right\rangle +\frac{1}{\sqrt{2}}e^{i\left(
b_{1}-b_{2}\right) t} \left\vert 10\right\rangle ,
\end{equation*}%
and the probability of being in the state $\left\vert 10\right\rangle $
after a time $t$ is 
\begin{equation*}
P_{-}
=\frac{1}{2}+\frac{1%
}{2}\cos \left[ 2\left( b_{1}-b_{2}\right) t\right] ,
\end{equation*}%
so the period is $\tau _{-}=\pi /\left\vert b_{1}-b_{2}\right\vert .$ \ As
above, we have 
\begin{equation*}
1/T_{2}^{\left( -\right) }\sim 4\left\langle \left[ \left( \delta b_{1}-\delta b_{2}\right) \right] ^{2}\right\rangle_\omega ,
\end{equation*}%
and the number of oscillations is%
\begin{equation*}
N_{-}=\frac{T_{2}^{\left( -\right) }}{\tau _{-}}\sim \frac{\left\vert
b_{1}-b_{2}\right\vert }{4\pi \left\langle \left[ \left(
\delta b_{1}- \delta b_{2}\right) \right] ^{2}\right\rangle_\omega }.
\end{equation*}

For perfectly correlated collective dephasing $\left( b_{1}=b_{2}\right) ,$ $N_{-}$
diverges, and again we have a DFS.

These two experiments suffice to identify the approximate DFS.  If $T_{2}^{+} >> T_{2}^{-}$ then we have the ``+'' subspace, 
while if If $T_{2}^{-} >> T_{2}^{+}$ then we have the ``-'' subspace,  This determination then fixes all the protocols that 
we will recommend below.  If there is strong noise that also flips the qubits, then (absent artificial symmetries) we do not expect to have even an approximate DFS. 

\ 
\begin{figure}[t]
\includegraphics[width=\linewidth]{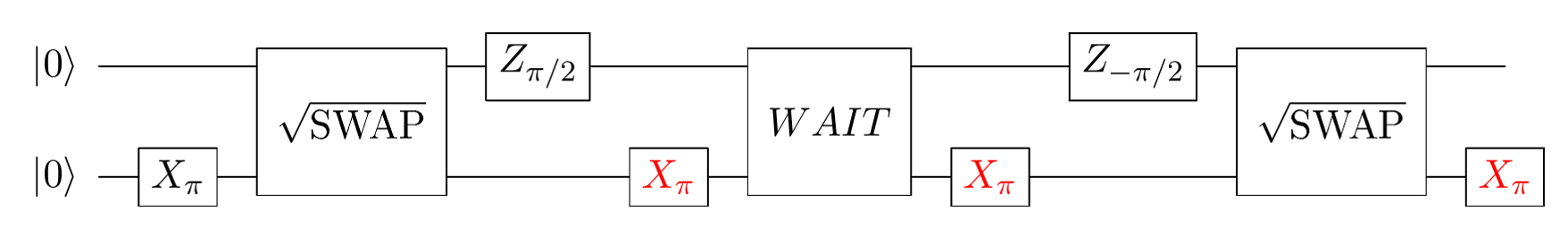}
\caption{A circuit for performing the Ramsey experiment described in Sec.
III. The final three $X$ rotations on the second qubit should be included to
prepare $\Psi _{-}$ and left out for $\Psi _{+}$.}
\label{fig:swseq}
\end{figure}

\subsection{Remarks}

There are several gate sequences that will perform the measurements of
correlated noise. \ A particularly simple set is shown in Fig. \ref%
{fig:swseq}. \ 

Once the determination of $T_{2}^{\left( +\right) }$ and $T_{2}^{\left(
-\right) }$ has been made, we extract the correlation function $S_{12}\left(
\omega \right) $ as follows. \ Since 
\begin{eqnarray*}
\frac{1}{T_{2}^{\left( \pm \right) }} &\sim &4\left\langle \left[ 
\left( \delta b_{1}\pm  \delta b_{2}\right) \right] ^{2}\right\rangle_\omega  \\
&=&4\left\langle \left( \delta b_{1}\right) ^{2}\pm 2\delta b_{1}\delta
b_{2}+\left( \delta b_{2}\right) ^{2}\right\rangle_\omega  
\end{eqnarray*}%
we deduce that
\begin{equation*}
\frac{1}{T_{2}^{\left( \pm \right) }}=\frac{1}{T_{2}^{\left( 1\right) }}+%
\frac{1}{T_{2}^{\left( 2\right) }}\pm \frac{8}{\hbar ^{2}}\lim_{\omega
\rightarrow 0}S_{12}\left( \omega \right) .
\end{equation*}%
This determines $S_{12}\left( \omega \right) ,$ which can also be thought of
as a quantity that breaks a sum rule on the $T_{2}$'s - the direction of the
breaking depending on whether the noise is correlated or anticorrelated. \
Of course this is only a rough relation. \ If there are differences in
frequency dependences of the various $S_{ij}\left( \omega \right) $ this can
modify the conclusions. \ There is of course nothing in this analysis that
limits it to a two-qubit system. \ In a many-qubit system, $S_{ij}\left(
\omega \right) $ can be measured in exactly the same way for all pairs. \ 

For ease of presentation we have stuck to the approximation that dephasing
depends only on the zero-frequency limit of the $S_{ij}$'s. \ Obviously this
is not the case for general frequency dependences, and the appropriate
modifications lead to substantial quantitative changes in the various $T_{2}$%
's \cite{Breuer2002}. \ However, the theory is
the same for $T_{2}^{\left( +\right) }$ and $T_{2}^{\left( -\right) }$ as it
is for the usual single-particle quantities $T_{2}^{\left( 1\right) }$ and $%
T_{2}^{\left( 2\right) },$ so the standard improved formulas can also be
applied to better determine the cross-correlation functions. \ Furthermore,
the noise spectroscopy experiments that have now become routine in
single-qubit experiments \cite{Yuge2011} to determine $%
S_{11}\left( \omega \right) $ can be done in exactly the same way to find $%
S_{12}\left( \omega \right) .$ \ The ``wait'' period in Fig. \ref{fig:swseq}
is modified to include a sequence of $X_1X_2$ gates corresponding to a frequency comb which decouples the system from noise at specific frequencies \cite{ Biercuk2009, Alvarez2011}. This can be varied to reconstruct the entire noise spectrum.     \

\section{Local Decoherence Measures}

With the noise correlations determined, how can we use the knowledge gained
to improve the performance of a quantum information processing device? \ DFS
theory offers a simple solution. We set the initial state of the computation to be in the DFS, and we design all subsequent unitary operations so that they never rotate the state out of the DFS. \ This procedure in principle eliminates decoherence. \ 

In the 2-qubit system with a 2-dimensional DFS we do not have this luxury.  The DFS is a two-dimensional subspace so no non-trivial quantum operations are possible. Furthermore, a realistic many-qubit system will not admit a perfect DFS. Still, we can hope to reduce decoherence for the two-qubit system, even if the DFS is only
approximate.  \ For a given processing task, many gate sequences are usually
possible. \ The idea is to choose the one most resistant to correlated noise. \ We achieve this by taking inspiration from the DFS procedure.

In the pure dephasing model we are considering, the only candidates for
perfect DFSs are $\Span\left\{ \left\vert 00\right\rangle ,\left\vert
11\right\rangle \right\} $ and $\Span\left\{ \left\vert 01\right\rangle
,\left\vert 10\right\rangle \right\} .$ If $T_{2}^{\left( +\right)
}/T_{2}^{\left( -\right) }$ is finite, then there is no DFS. \ Nevertheless,
if $T_{2}^{\left( +\right) }/T_{2}^{\left( -\right) }<1,$ then $\Span\left\{
\left\vert 01\right\rangle ,\left\vert 10\right\rangle \right\} $ is the
``good'' subspace and if $T_{2}^{\left( +\right) }/T_{2}^{\left( -\right) }>1,$
then $\Span\left\{ \left\vert 11\right\rangle ,\left\vert 00\right\rangle
\right\} $ is the ``good'' subspace. \ As might be expected from symmetry, our
conclusions are equally valid for the two cases. \ 

Since the subspaces are good but not perfect, this picture suggests the idea
of defining a measure of decoherence for every point in the Hilbert space
when noise correlations are present. \ We call these ``local'' decoherence
measures, ``local'' here referring to Hilbert space, not real space.

We define two such measures. \ 

1. The first is a geometric measure, called $d_{g}\left( \left\vert \psi
\right\rangle \right) ,$ where $\left\vert \psi \right\rangle $ is any
vector in the 4-dimensional 2-qubit space. \ Let $B$ be an orthonormal basis
for the DFS. \ The projection operator onto the DFS is 
\begin{equation*}
P=\sum_{\left\vert \phi \right\rangle \in B}\left\vert \phi \right\rangle
\left\langle \phi \right\vert ,
\end{equation*}%
and then 
\begin{equation*}
d_{g}=|(I-P)\psi |^{2}
\end{equation*}%
returns the square of the perpendicular Hilbert space distance from $\psi $
to the DFS. \ This gives a very simple geometric picture of the decoherence
rate of the state as depending only on the distance to the DFS. \ Clearly $%
d_{g}\leq 1$ and $d_{g}=0$ for $\left\vert \psi \right\rangle $ in the DFS.
\ 

2. The second decoherence measure, called $d_{c}$ $\left( \left\vert \psi
\right\rangle \right) ,$ is obtained by studying the purity $\gamma $ of a
density matrix $\rho $ of the 2-qubit system, defined as $\gamma =$Tr$\left(
\rho ^{2}\right) $. $\gamma =1$ for a pure state since then $\gamma =$Tr$%
\left( \rho ^{2}\right) =$Tr$\left( \rho \right) =1.$ \ For the completely
mixed state $\rho =I/D,$ where $D$ is the dimension of the Hilbert space we
find $\gamma =1/D$. \ Our interest is in the case $D=4.$ \ To use $\gamma $
to form a local measure of decoherence in the Hilbert space we imagine
initializing the system at time $t=0$ in the state $\rho \left( 0\right)
=\rho _{0}=\left\vert \psi \right\rangle \left\langle \psi \right\vert $ so
that $\gamma \left( t=0\right) =1$ and then watching $\gamma $ decrease with time
under the influence of the noise Hamiltonian $H_{n}$. \  Denote averages of 
$\cdot $ over noise realizations by $\left[ \cdot \right] _{av}$. \  In this
case, $d\rho /dt=\rho ^{\prime }=-i\hbar \lbrack \rho ,H_{n}]_{av}$ . \ We
are only interested in the short-time behavior of $\gamma $ so we get 
\begin{align*}
\gamma (\delta t)& =\text{Tr}\left[ \rho (\delta t)\right] ^{2} \\
& \approx 1-\text{Tr}\left[ \delta t\rho _{0}^{\prime }+\frac{1}{2}\delta
t^{2}\rho _{0}^{\prime \prime }\right] ^{2} \\
& =1-\frac{\delta t^{2}}{2}\text{Tr}\rho _{0}\rho _{0}^{\prime \prime } \\
& =1-\delta t^{2}\text{Tr}\left[ \rho _{0}^{2}H_{n}^{2}-\rho _{0}H_{n}\rho
_{0}H_{n}\right] _{av}
\end{align*}
We identify 
\begin{equation*}
d_{c}(\left\vert \psi \right\rangle )=Tr\left[ \rho _{0}^{2}H_{n}^{2}-\rho
_{0}H_{n}\rho _{0}H_{n}\right] _{av}
\end{equation*}
as a measure of decoherence that describes how susceptible the pure state $%
\rho (t)$ is to mixing by the noise. \ Once again, $\rho _{0}=\left\vert
\psi \right\rangle \left\langle \psi \right\vert $ so $d_{c}$ is a
decoherence measure associated with a point in the Hilbert space of the
computer. \ $d_{c}=0$ if $\left\vert \psi \right\rangle $ is in the DFS
since $H_{n}$ acts as a constant operator in the DFS and $\left[ \rho
_{0},H_{n}\right] =0.$ \ Unlike $d_{g},$ however, there is no upper bound on 
$d_{c}$, and $d_{c}$ has no natural normalization. \ However, it is only used for comparison of circuits, so this is not a severe drawback.    

The two measures differ considerably in their generality. \ $d_{g}$ relies
only on the identification of a DFS and can be thought of as an extension of
the DFS concept. \ By contrast, to compute $d_{c}$ one needs only the noise
Hamiltonian. \ $d_{g}$ is simple to compute and to visualize. \ But $d_{c}$
gives a more complete picture of the decoherence. \ It is possible that the
decoherence is not even a monotonic function of the distance from the
approximate DFS. \ $d_{g}$ obviously does not capture this possibility. \
Finally, $d_{c}$ can clearly be computed also for mixed states, while $d_{g}$
cannot be, at least by the above definition.

In the course of a quantum information process, an ideal computer remains in
a pure state $\left\vert \psi \left( t\right) \right\rangle $ that traverses
a path in Hilbert space from the initial state $\left\vert \psi \left(
t=0\right) \right\rangle $ to the desired final state $\left\vert \psi
\left( t=t_{f}\right) \right\rangle $ that encodes the answer to the
computation or other process. \ Given this trajectory we can also compute $%
d_{g}\left( t\right) $ and $d_{c}\left( t\right) .$ \ If these quantities
are big on average over the interval $0\leq t\leq t_{f},$ then we expect
poor fidelity in the result. \ Of course there is a choice of gate sequences
(actually an infinite number) that will take the computer from $\left\vert
\psi \left( t=0\right) \right\rangle $ to $\left\vert \psi \left(
t=t_{f}\right) \right\rangle $. \ The choice is usually determined by
brevity and experimental constraints.

The central contention of this paper is that one should also
take into account the minimization of decoherence. \ A gate sequence that
minimizes $d_{g}\left( t\right) $ and/or $d_{c}\left( t\right) $ in the presence of correlated noise should be preferred. \ Of course for this small system the fidelity itself can easily be computed and used to minimize the decoherence. \ However, it is often difficult to understand purely numerical calculations of the infidelity, and the use of $d_{g}$ and $d_{c}$ gives physical insight and, as we shall see, can also suggest generalizations to larger systems.

\section{Results}
We test these ideas on two quantum information processing tasks that can be carried out in two-qubit systems, the Deutsch-Jozsa algorithm and Bell-state preparation. \ These choices were motivated mainly by the fact that they have been carried out successfully in recent experiments \cite{Watson2018} so we can use sequences that have actually been shown to be successful.  Note that the Bell-state preparation is part of the circuit required for the measurements described in Sec. III.

\subsection{Noise Model}

During the course of the tasks the system is subjected to quasi-static noise with $\vec{%
\delta b} \equiv (\delta b_1, \delta b_2)^T$ sampled from a bivariate Gaussian
distribution with density 
\begin{equation*}
f(\delta b_1, \delta b_2) = \frac{1}{2 \pi \sqrt{ \det \Sigma }} \exp \left(
- \frac{1}{2}\vec{\delta b}^T \Sigma^{-1} \vec{\delta b} \right).
\end{equation*}
The model assumes zero mean (any deviation from this can be absorbed into the static Hamiltonian $H_0$) and covariance 
\begin{equation*}
\Sigma = \left(%
\begin{array}{cc}
\sigma_1^2 & c \sigma_1 \sigma_2 \\ 
c \sigma_1 \sigma_2 & \sigma_2^2%
\end{array}
\right)
\end{equation*}
where $\sigma_{1,2}$ is the noise strength at qubits $1$ and $2$ and and $c$
is their statistical correlation. \ We begin with a simple model in
which there is complete correlation of the noise: $\delta b_{1}\left( t\right) \propto \delta
b_{2}\left( t\right)$.  The strength of the noise may be different on the two qubits.  This is quantified by a qubit asymmetry $r = \sigma_1/\sigma_2$, the ratio of the width of the field distribution on qubit $1$ to that
on qubit $2$. For $r=1$ we are in the ``-'' subspace.  Decoherence is simulated by averaging the dynamics over many realizations of the noise.  The number of realizations is determined by examining the convergence of the computed quantities as the number increases.  We used a convergence criterion of $2\%$, which was typically achieved after averaging over about 1000 realizations. 
\begin{figure}[h]
\includegraphics[width=1.0%
\linewidth]{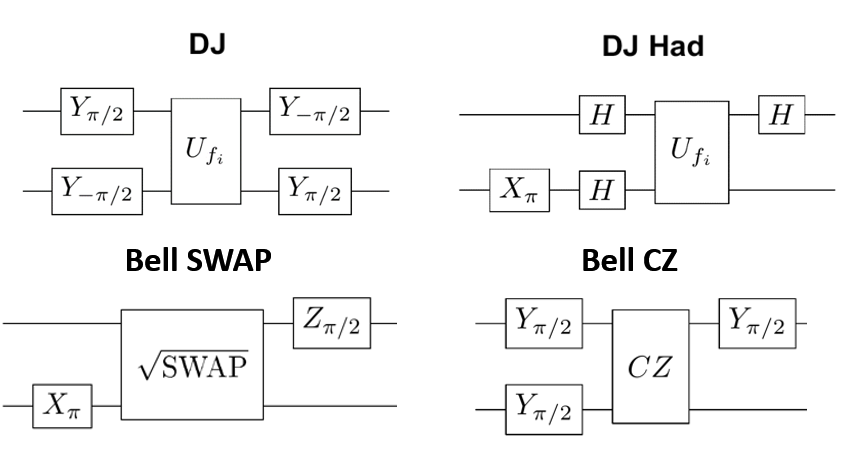}
\caption{Gate sequences for performing common two-qubit information processing tasks.  Top left (right) is the Y-gate (Hadamard gate) circuit for performing the Deutsch-Josza algorithm.  Bottom left (right) is the $\sqrt{SWAP}$ (CZ) circuit for preparing a Bell state.}
\label{fig:circuits}
\end{figure}
\subsection{Deutsch-Jozsa Algorithm}
As the first example, we consider two gate sequences for performing the Deutsch-Jozsa algorithm with a quantum oracle that encodes a balanced function. Following \cite{Watson2018}, $U_{f_i} = CNOT$ is implemented via exchange interaction and single qubit rotations.\  Two circuits that perform this algorithm are given in Fig. \ref{fig:circuits}. Even though the initial and final states are the same, the sequences differ very substantially, particularly in their one-qubit gates.  

\begin{figure}[tbp]
\includegraphics[width=1.0%
\linewidth]{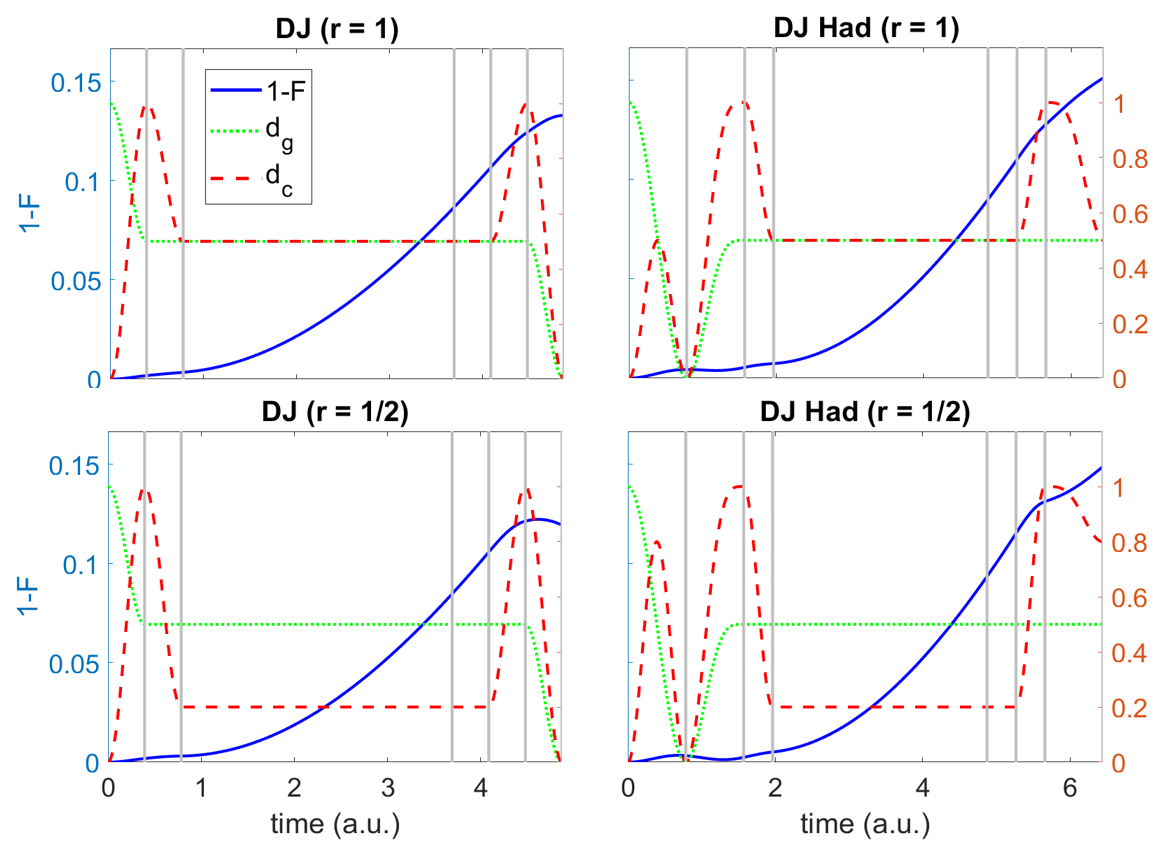}
\caption{Two different circuits (Fig. \ref{fig:circuits}) for performing the Deutsch-Jozsa
algorithm are compared using infidelity along with $d_g(t)$ an $d_c(t)$ for the
function $U_{f_i} = CNOT$. The circuits are subject to perfectly correlated quasistatic Gaussian dephasing noise where the ratio between the noise strength at the two qubit locations is $r$. Gray lines divide the time axis into intervals corresponding to each gate in the circuit.}
\label{fig:djcc}
\end{figure}

We calculate the ideal unitary evolution for the noise-free system, easily obtained from the gate sequences. \ We also compute the non-unitary dynamics of each circuit when it is subjected to noise.  The results are summarized in Fig. \ref{fig:djcc}. Comparison of the results of the two calculation allows us to plot the infidelity ($1-F$, where $F$ is the fidelity) as a function of time. \ For the computation of $d_{g}(t) $ and $d_{c}(t)$ we need only the noise-free state.  $d_g(t), d_c(t)$ and the infidelity are all plotted as a function of time in arbitrary units for two different values of $r$.  The relative times for each gate are taken from Ref. \cite{Watson2018}.  The success of the decoherence measures should be judged by the extent to which they resemble the time derivative of $1-F$.  

We focus first on the left panel of the Fig \ref{fig:djcc}, in which the ``Y-gate'' circuit is analyzed. 
One first notes that although $d_g(t)$ and $d_c(t)$  track each other for a substantial portion of the evolution, there is also quite a bit of disagreement between them reflecting the fact that decoherence is not just a matter of distance from the DFS. In part this is because $d_g(t)$ is not sensitive to varying $r$: indeed its construction assumes an $r = 1$ DFS.  This is true even though we chose a quite simple noise model.  The difference between $d_g(t)$ and $d_c(t)$ allows us to distinguish certain ways in which one is better than the other.  For example, at short times $1-F$ rises quadratically with time.  This is in disagreement with the geometric measure, since $d_g(0)$ is finite.  The purity measure, which is linear in $t$ at small times, does better.  On the other hand, both measures capture the leveling-off of the infidelity at the end of the interval.  In between, the main difference in $d_g(t)$ and $d_c(t)$ is the two bumps in $d_c(t)$.  This is reflected only to a very small extent in $1-F$.  

The results for the two values of $r$ are rather similar.  There is one interesting difference at around $t=1.5$, where $d_c(t)$ captures the momentary leveling-off in the infidelity better.  Comparing $r=1$ (top) with $r=1/2$ we see that $d_g$ is not affected by $r$, whereas $d_c$ and $1-F$ are somewhat reduced.  Again, $d_c$ seems to be the slightly better measure.   

The right panel shows the same analysis for the ``H-gate'' circuit.  This circuit has an anomalous region, near $t=2$, where the infidelity actually decreases with time.  One can trace this behavior back to an echo effect provided by the X and H gates.  These subtleties are not captured by $d_g(t)$ or $d_c(t)$, which are of course both non-negative.  Apart from this, the virtues and deficiencies in $d_g(t)$ and $d_c(t)$ are as in the other circuit.  Note that both predict the increase in the infidelity at the end of the time interval.    

As for using $d_g(t)$ and $d_c(t)$ to decide between the two circuits, the anomalous echoing effect clearly reduces the usefulness of the two measures.  Overall, both $d_g(t)$ and $d_c(t)$ are larger for the ``Y-gate'' circuit, but the final $1-F$ for the two circuits is actually about the same.

\begin{figure}[t]
	\includegraphics[width=\linewidth]{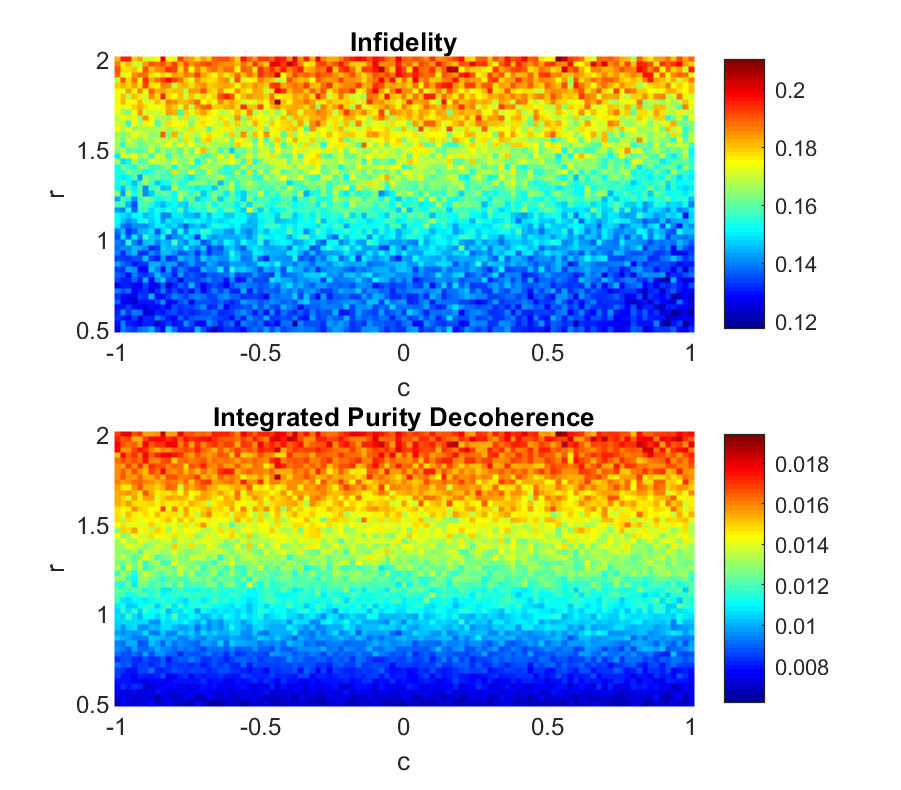}
	\caption{The final infidelity and integrated purity decoherence $d_c$ for
		the Y-gate Deutsch-Jozsa gate sequence with $U_{f_i} = CNOT$.}
	\label{fig:fidmapdj3}
\end{figure}

In Fig.\ref{fig:fidmapdj3} we compare the final $1-F$ and the time integral of $d_c(t)$  for a large range of values of the asymmetry $r$ and the correlation $c$ for the Y-gate circuit.  In this circuit there is no anomalous behavior of $1-F$.  It seems that when this is the case, then the integral of $d_c(t)$ is indeed a good predictor of fidelity for correlated ($c=1$), uncorrelated ($c=0$) and anti-correlated ($c=-1$) noise.  This is true even when the noise is much stronger on one of the qubits.  

\subsection{Bell-state Preparation}

\begin{figure}[h]
\includegraphics[width=%
\linewidth]{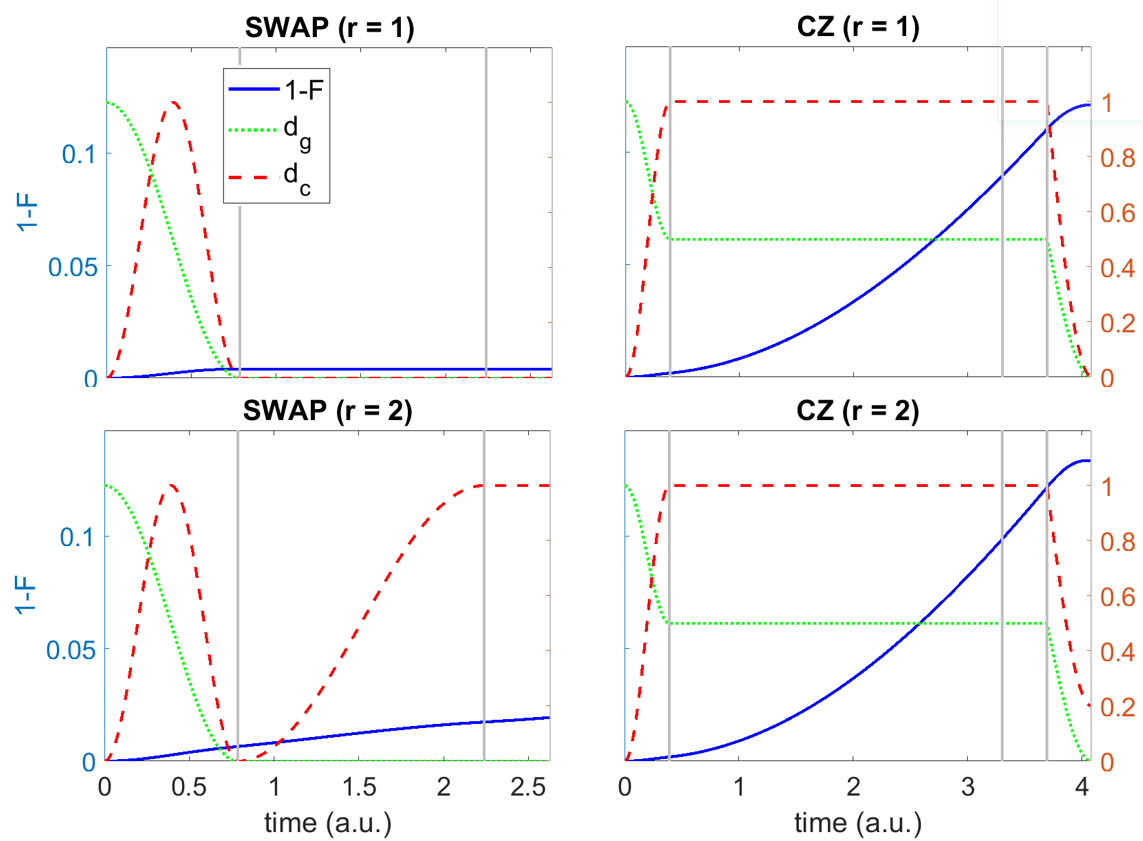}
\caption{Comparison of two circuits for preparing the state $\Psi =( 
\ket{\downarrow \uparrow}+\ket{\uparrow \downarrow})/\protect\sqrt{2}$ to
perform one of the measurements described in Sec. II. The left panels are for the Bell SWAP circuit and the right panels are for the Bell CZ circuit.  The top panels are for $r=1$ so the noise strength is the same on the two qubits.  The bottom panels are for $r=2$ so the noise strength is stronger on qubit 1.  The solid blue, dotted green, and red dashed lines are respectively the state infidelity $1-F$, the DFS projection metric $d_g$, and the purity-based decoherence $d_c$ for each circuit as a function of time.  The system is subject to quasistatic Gaussian dephasing noise. $d_{g}(t)$ and  $d_{c}(t)$ are scaled and
overlaid against the infidelity to illustrate how they capture dephasing
effects. Gray lines divide the time axis into intervals corresponding to
each gate in the circuit. }
\label{fig:ramseycircuitcomparison}
\end{figure}

The circuits start from an initial state $\left\vert 00\right\rangle $ and end in
the Bell state $\left( \left\vert 01\right\rangle +\left\vert
10\right\rangle \right) /\sqrt{2}.$ The first circuit we call the $\sqrt{SWAP}$ circuit and the second is the $CZ$ circuit.  The names reflect the fact that the main difference between the two circuits is the nature of the entangling gate.   $d_{g}(t)$, $d_{c}(t)$, and $1-F(t)$ are plotted in Fig. \ref{fig:ramseycircuitcomparison}. 

The comparison of the two circuits is more straightforward here, since there is no anomalous behavior in $1-F(t)$.  \ In the $CZ$ circuit, the time integrals of $d_g(t)$ and $d_c(t)$ are both clearly bigger than in the  $\sqrt{SWAP}$ circuit.  Both measures predict that the final $1-F$ should be bigger for the  $CZ$ circuit, and indeed it is.  It is also true that the shape of $1-F(t)$ resembles the integral of $d_c(t)$. 

Passing to the $r$-dependence of the $\sqrt{SWAP}$ circuit, we note that at later times $t>2$, $d_{c}\left( t\right) $ shows significant differences between the $r=1$ and $r=2$ cases. \ This is faithfully reflected in the higher final infidelity for $r=2$. \
We note once more that near $t=0,$ d$_{c}\left( t\right) $ is linear in time, while $d_{g}( t=0)$ is finite and $1-F$ is quadratic.  Thus $d_{c}\left( t\right)$ always seems to be superior to $d_{g}\left( t\right)$ when $1-F$ is small.

\begin{figure}[t]
\includegraphics[width=\linewidth]{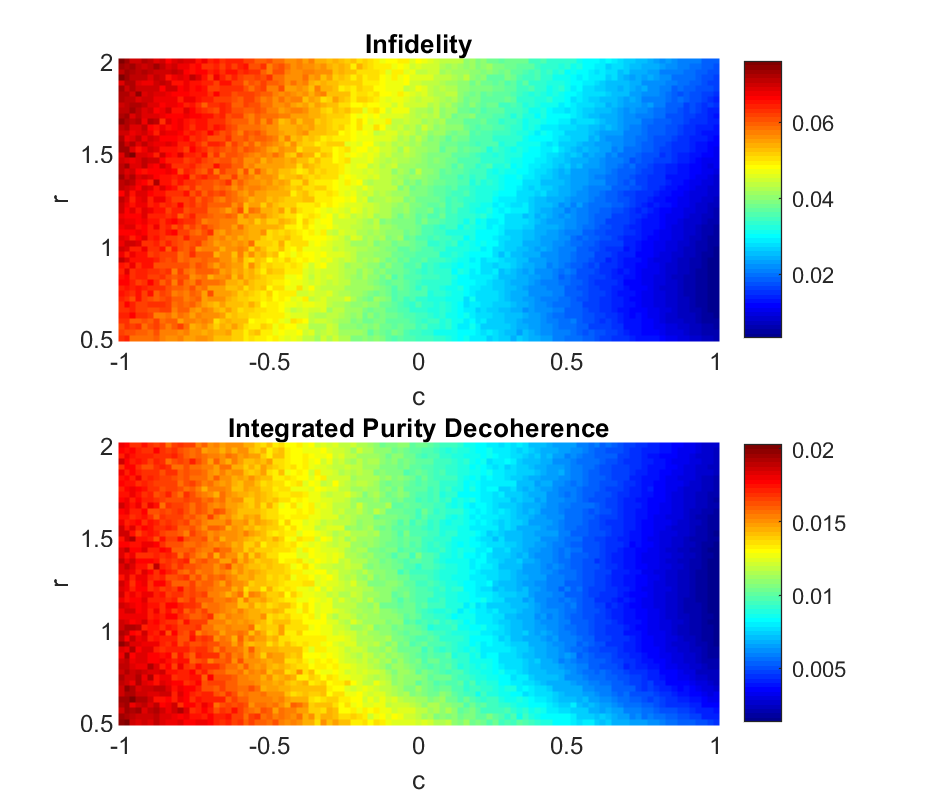}
\caption{The final infidelity and averaged purity decoherence $d_c$ for the $%
\protect\sqrt{\text{SWAP}}$ Bell-state circuit. As explained in the text, the purity
decoherence serves as a good predictor for the derivative of the fidelity so the
integral, or average, over the gate sequence shares the same features as the
infidelity. }
\label{fig:fidmapramsey}
\end{figure}

\begin{figure}
	\includegraphics[width=0.7\linewidth]{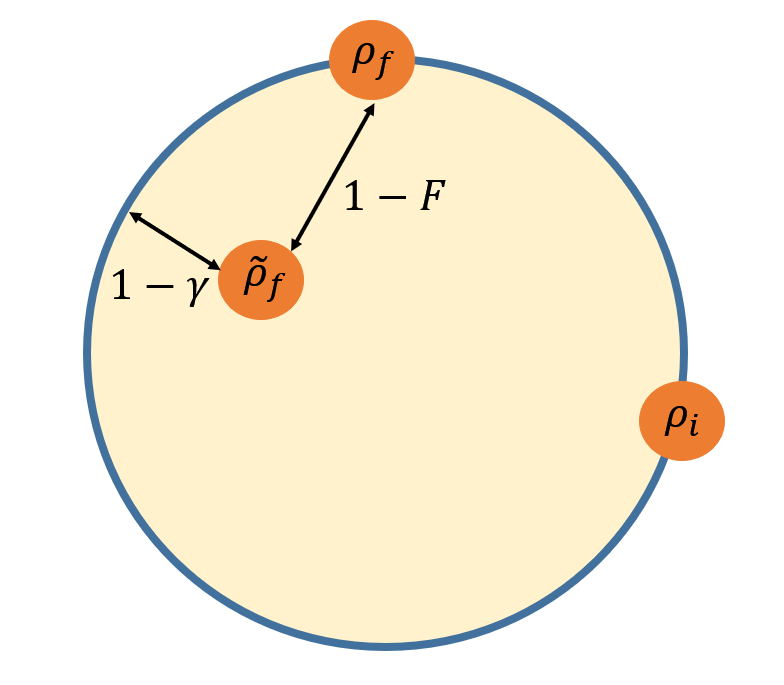}
	\caption{Schematic projection of the 2-qubit state space $R$ as a ball with the boundary containing the pure state submanifold. The pure initial ($\rho_i$) and error-free final ($\rho_f$) states of an algorithm, and the partially decohered result of running that algorithm in a noisy environment are labeled ($\widetilde{\rho}_f$). State fidelity $F$ provides a measure of closeness between the two states $\rho_f$ and $\widetilde{\rho}_f$, while the impurity $1 - \gamma$ only represents closeness to the pure state subspace.  }
	\label{fig:2dstatespace}
\end{figure}

In Fig. \ref{fig:fidmapramsey} we again plot the the final $1-F$ for a range of $r$ and $c$ and compare it to the time integral of $d_c(t)$ for the $\sqrt{\text{SWAP}}$
Bell-state circuit.  In contrast to the Deutsch-Jozsa Y-gate circuit, there are substantial differences between the two quantities. We see that the impurity is roughly independent of the asymmetry $r$, both qubits contributing roughly equally.  In contrast, the fidelity depends more strongly on the asymmetry.  This comes from the asymmetry of the circuit itself, specifically that there is an X-gate applied to qubit 2 but not to qubit 1.  The X-gate echoes away the decoherence created by noise on qubit 2 but not that on qubit 1.  Thus, once more we see that echo effects can reduce the information supplied by local decoherence measures.  This indicates a subtle but important drawback to the use of $d_c(t)$ as a circuit quality measure, which stems ultimately from the difference between purity and fidelity. 
Consider the enlarged state space of the 2-qubit circuit as the real 15-dimensional Hilbert space of density matrices (actually a compact subset of this space when positivity constraints are added.)  Each density matrix is a point in the space.  The pure states live in a 6-dimensional submanifold.  $1-F$ is a measure of the distance in this space from the desired final state to the actual one: we may think of it as the length of the difference vector.  The desired final state is a pure state.  The integral of $d_c(t)$, however, only provides one component of the difference vector - essentially the vector that is perpendicular to the subspace of pure states.  There is also a component of the difference vector parallel to the subspace.  The vectors are shown schematically in Fig. \ref{fig:2dstatespace}.

\section{Extension to Many Qubits}

As noted above, if only a few qubits are involved and the noise model is known, the best strategy for deciding on a gate sequence is simply to calculate the infidelities. \ As
the number of qubits increases, the length of this computation increases
exponentially, and it soon becomes impractical. \ The same holds for the
computation of $d_{g}\left( t\right) $ and $d_{c}\left( t\right) ,$ since
they depend on the many-body wavefunction. The question is whether we can
use the physical insight gained for few-qubit systems to give a meaningful
presciption for ``scoring'' long gate sequences in a multi-qubit computer.\ \
\ 

We propose that this is indeed possible, based on a picture of the errors that occur during the computation as steps in a random walk, the walk taking place in a moving frame generated by the algorithm.

The density matrix space $R$ for an $n$-qubit computer has real dimension $4^{n}-1$.  The Hilbert Schmidt inner product on $R$ defines what is essentially a Euclidean distance metric.  An error-free computation with circuit depth $k$ consists of a sequence of $k$ points in this space, labeled $\rho_t$, $t=0,1,2...,k$.  The states $\rho_t$ are all pure, so this ideal evolution takes place in the $2^{n+1}-2$-dimensional submanifold of pure states. This non-random evolution moves by large distances at each step and defines a moving frame in $R$. In this moving frame the errors define a random walk, assuming that there is no correlation between the sequence defined by the algorithm and the local
decoherence.   

\ Taking a clue from the definition of $d_{g}$,
decoherence will be worse if the random walk increases the perpendicular
distance of $\rho_t$ to the DFS. \ If we could compute a
probability distribution $P_{\perp }$ for the perpendicular component of the
steps in the walk then the expected value of the perpendicular distance after $t$ gates would be $\langle d_{t}^{2}\rangle ^{1/2}=\sqrt{t}L$, where $L$ is the rms step length computed using $P_{\perp }.$ \ We may
think of different gate sequences as having different $P_{\perp }.$

We therefore assign a \textquotedblleft perpendicular
step size\textquotedblright\ to every one or two qubit gate $G.$ Since the walk takes place close to the manifold of pure states, \ We can use the usual Hilbert space  $\mathcal{H}$ and we arrange the basis of $\mathcal{H}$ so that the $D$ basis vectors of the approximate DFS, which we call $\mathcal{C}$, come first, and the $2^{n}-D$ vectors of the orthogonal complement $\mathcal{C}_{\perp }$ come second. We then partition the unitary matrix of $G$ into four
sectors: 
\begin{equation*}
G=\left( 
\begin{array}{cc}
G_{\parallel } & M \\ 
M^{\prime } & G_{\perp }%
\end{array}
\right) ,
\end{equation*}%
where $G_{\parallel }$ and $G_{\perp }$ move the state around $\mathcal{C}$
and $\mathcal{C}_{\perp }$ respectively. \ $M$ and $M^{\prime }$ move weight
between $G_{\parallel }$ and $G_{\perp }.$ \ A ``good'' gate has $M=M^{\prime
}=0$ since a state in $C$ undergoes no perpendicular motion and therefore remains in $%
\mathcal{C}$ and a state in $\mathcal{C}_{\perp }$ undergoes random motion
that is not biased in the perpendicular direction. \ $M$ and $M^{\prime }$
give perpendicular motion and if they are large then $G$ is
a bad gate, i.e., one that we expect will increase the decoherence to which
the state is exposed. \ To quantify this we use the unitarity of $G$ and
index the blocks of $G$ as follows. \ The upper left corner $G_{\parallel }$
is indexed by $s_{G_{\parallel }}=(i,j)$ such that $1\leq i\leq D$ and $%
1\leq j,\leq D$; the lower right corner $G_{\perp }$ is indexed by $%
s_{G_{\perp }}=(i,j)$ such that $D+1\leq i\leq D_{H}$ and $D+1\leq j\leq
D_{H}$; the upper right corner $M$ is indexed by $s_{M}=(i,j)$ such that $%
1\leq i\leq D$ and $D+1\leq j\leq D_{H};$ the lower left corner $M^{\prime }$
is indexed by $s_{M^{\prime }}=(i,j)$ such that $D+1\leq i\leq D_{H}$ and $%
1\leq j\leq D.$ 

We then define 
\begin{equation*}
B(G)=\frac{1}{4}\left( \sum_{(i,j)\in s_{M}}|M_{ij}|^{2}+\sum_{(i,j)\in
s_{M'}}|M_{ij}^{\prime }|^{2}\right) ^{1/2},
\end{equation*}%
which is just a Euclidean measure of the size of the off-diagonal blocks. \
This may be thought of as the ``badness'' of $G.$ \ The unitarity of $G$ and
the fact that one- and two-qubit gates are nearly diagonal imply that $%
0\leq B\leq 1.$ \ Table \ref{tab:gatescores} gives the badness of several
common gates. A circuit $A$ then may be thought of as a random walk with $%
N_{A}$ gates and the analog of the integral of $d_{g}$ over the total
circuit is 
\begin{equation*}
d_{A}=\sqrt{\sum_{G}B^{2}\left( G\right) }.
\end{equation*}

We expect the infidelity to be roughly proportional to this quantity.
\begin{table}[t!]
\vspace{10pt}
\begin{minipage}[t]{0.33\linewidth}
		\begin{tabular}[t]{l r}
			\hline\hline 
			$G$ & $B(G)$  \\
			\hline
			$X$ & $1/2$\\
			$Y$ & $1/2$ \\
			$Z$ & $0$ \\
			$H$ & ${2^{-5/4}}$ \\
			CNOT & $2^{-3/2}$ \\
			\hline\hline 
		\end{tabular}
	\end{minipage}
\begin{minipage}[t]{0.33\linewidth}
		\begin{tabular}[t]{l r r} 
			\hline\hline
			$C$ & $d_A$ & $1-F$ \\
			\hline
			Bell SWAP & $0.500$ & $0.004$\\
			Bell CZ & $0.728$ &$0.011$ \\
			DJ  & $0.912$  &$0.019$\\
			DJ Had & $0.951$ &$0.034$ \\
			\hline\hline
		\end{tabular}
	\end{minipage}
\caption{On the left, badness $B(G)$ for some typical one and two qubit gates $G$. The right table shows $d_A$ for a few of the circuits studied above, demonstrating the correspondence between $d_A$ and final state infidelity. The
calculation assumes the existence of an approximate DFS spanned by $%
\{\ket{\uparrow \downarrow },\ket{\downarrow \uparrow }\}$.}
\label{tab:gatescores}
\end{table}
The key point is that we never need to compute any wavefunctions or other
many-body quantities. \ Each one- and two-qubit gate has only a small number
of off-diagonal elements, so the computation of $B$ is efficient - in fact
it is very fast. \ Thus we may score different gate sequences and choose the
right one for our multi-qubit computation without prohibitive overhead.

We now propose a protocol for improving the performance of a many-qubit computer.  The first step is a calibration phase, in which the experiments given in Sec. III are performed for each pair of qubits.  This gives all two-point correlation functions.  Each pair of qubits is assigned a ``+'' sign or a ``-'' sign according to the results for $T_{2}^{+}$ and $T_{2}^{-}$.  This then determines an approximate DFS for each pair.  In the second step  every one- and two-qubit gate $G$ is assigned a score $B(G)$ and each candidate circuit $A$ is assigned a score $d_{A}$.  Finally, the circuit with mininum $d_{A}$ is chosen.

Our model system gives numerical evidence for the validity of this idea, since it can process tasks with distinct circuits. These are the same circuits for Bell state preparations and the Deutsch-Jozsa algorithm from Sec V.  Table VI.I shows some individual gate scores, and then $d_A$ and $1-F$ for 4 circuits.  One sees that $1-F$ and $d_A$ are monotonically related, but not strictly proportional. This result is encouraging, but it is obtained for a very small system.  Further work on larger systems will be needed to confirm the basic concepts and to refine the protocol.    

\section{Conclusion}

We have presented a method to measure spatial noise correlations in quantum information processors, focusing on those correlations that are most important for error mitigation. \ It was formulated for a 2-qubit machine, but it is clearly also immediately applicable to any 2-point correlation in a machine of arbitrary size.  This information is sufficient to identify approximate DFSs, which in turn informs the design of gate sequences. This is done by identifying decoherence measures that are local in Hilbert space, and using sequences that avoid regions where these measures are high.  These measures can only be computed in few-qubit systems, which limits their usefulness.  However, they point the way to a method that assigns scores to individual gates even in many-qubit systems.  By means of a picture of errors generating a random walk in the state space, we can give a score to any candidate circuit.  A circuit with a low score will be more resistant to correlated noise.  This is confirmed by numerical calculation on a two-qubit system.

The method clearly does not offer a complete picture of the situation.  

We found in particular that some circuits have echo effects that actually increase the fidelity (at least for a short time).  Our decoherence measures do not capture this, though it is unclear whether this is ever a large effect.  

It is also unlikely that the method is very useful for very general error models.  We considered only dephasing noise.  It is easy to produce 2-dimensional DFSs for this type of noise.  If other noise that, for example, flips spins, is added, these DFSs disappear immediately. That does not prevent us from defining the decoherence measure $d_c$, which we found to be the most useful one, but if it varies little as we move around the space, it loses its power to distinguish different circuits.

The many-qubit method is however well-designed to be used in conjunction with error correction.  It assumes that the system stays reasonably close to the pure state manifold, meaning that its usefulness degrades as $k$, the circuit depth, increases.  However, if $k$ instead represents the number of gates that are performed between each error-correction cycle, $k$ can optimized to take advantage of our method. $k$ will be larger for better circuits.   
\begin{acknowledgments}
We acknowledge J.M. Boter, X. Xue, H. Ekmel Ercan, T.F. Watson, Joydip Ghosh, L.M.K. Vandersypen, Mark Friesen, M.A. Eriksson and S.N. Coppersmith for helpful discussions during the formative stages of this work, particularly towards Sec. III. This research was sponsored in part by the Army Research Office (ARO) under Grant Numbers W911NF-17-1-0274.The views and conclusions contained in this document are those of the authors and should not be interpreted as representing the official policies, either expressed or implied, of the Army Research Office (ARO), or the U.S. Government. The U.S. Government is authorized to reproduce and distribute reprints for Government purposes notwithstanding any copyright notation herein. 
\end{acknowledgments}

\bibliographystyle{unsrt}
\bibliography{bibliography}

\begin{thebibliography}{10}

\bibitem{Preskill2013}
John Preskill.
\newblock Sufficient condition on noise correlations for scalable quantum
  computing.
\newblock {\em Quantum Info. Comput.}, 13(3-4):181--194, March 2013.

\bibitem{Preskill1998}
John Preskill.
\newblock {\em Fault-Tolerant Quantum Computation}, pages 213--269.
\newblock World Scientific, 1998.

\bibitem{Bombin2016}
H\'ector Bomb\'{\i}n.
\newblock Resilience to time-correlated noise in quantum computation.
\newblock {\em Phys. Rev. X}, 6:041034, Nov 2016.

\bibitem{Lidar1998}
D.~A. Lidar, I.~L. Chuang, and K.~B. Whaley.
\newblock Decoherence-free subspaces for quantum computation.
\newblock {\em Phys. Rev. Lett.}, 81:2594--2597, Sep 1998.

\bibitem{Duan1998}
Lu-Ming Duan and Guang-Can Guo.
\newblock Reducing decoherence in quantum-computer memory with all quantum bits
  coupling to the same environment.
\newblock {\em Phys. Rev. A}, 57:737--741, Feb 1998.

\bibitem{Lidar2001}
Daniel~A. Lidar, Dave Bacon, Julia Kempe, and K.~B. Whaley.
\newblock Decoherence-free subspaces for multiple-qubit errors. i.
  characterization.
\newblock {\em Phys. Rev. A}, 63:022306, Jan 2001.

\bibitem{Lidar2001_2}
Daniel~A. Lidar, Dave Bacon, Julia Kempe, and K.~B. Whaley.
\newblock Decoherence-free subspaces for multiple-qubit errors. ii. universal,
  fault-tolerant quantum computation.
\newblock {\em Phys. Rev. A}, 63:022307, Jan 2001.

\bibitem{Temme2017}
Kristan Temme, Sergey Bravyi, and Jay~M. Gambetta.
\newblock Error mitigation for short-depth quantum circuits.
\newblock {\em Phys. Rev. Lett.}, 119:180509, Nov 2017.

\bibitem{Watson2018}
T.~F. Watson, S.~G.~J. Philips, E.~Kawakami, D.~R. Ward, P.~Scarlino,
  M.~Veldhorst, D.~E. Savage, M.~G. Lagally, Mark Friesen, S.~N. Coppersmith,
  M.~A. Eriksson, and L.~M.~K. Vandersypen.
\newblock A programmable two-qubit quantum processor in silicon.
\newblock {\em Nature}, 555:633 EP --, Feb 2018.

\bibitem{Friesen2017}
Mark Friesen, Joydip Ghosh, M.~A. Eriksson, and S.~N. Coppersmith.
\newblock A decoherence-free subspace in a charge quadrupole qubit.
\newblock {\em Nature Communications}, 8:15923 EP --, Jun 2017.
\newblock Article.

\bibitem{Zwanenburg2013}
Floris~A. Zwanenburg, Andrew~S. Dzurak, Andrea Morello, Michelle~Y. Simmons,
  Lloyd C.~L. Hollenberg, Gerhard Klimeck, Sven Rogge, Susan~N. Coppersmith,
  and Mark~A. Eriksson.
\newblock Silicon quantum electronics.
\newblock {\em Rev. Mod. Phys.}, 85:961--1019, Jul 2013.

\bibitem{Biercuk2009}
Michael~J. Biercuk, Hermann Uys, Aaron~P. VanDevender, Nobuyasu Shiga, Wayne~M.
  Itano, and John~J. Bollinger.
\newblock Optimized dynamical decoupling in a model quantum memory.
\newblock {\em Nature}, 458:996 EP --, Apr 2009.

\bibitem{Alvarez2011}
Gonzalo~A. \'Alvarez and Dieter Suter.
\newblock Measuring the spectrum of colored noise by dynamical decoupling.
\newblock {\em Phys. Rev. Lett.}, 107:230501, Nov 2011.

\bibitem{Yuge2011}
Tatsuro Yuge, Susumu Sasaki, and Yoshiro Hirayama.
\newblock Measurement of the noise spectrum using a multiple-pulse sequence.
\newblock {\em Phys. Rev. Lett.}, 107:170504, Oct 2011.

\bibitem{Sza2016}
Piotr Sza\ifmmode~\acute{n}\else\'{n}\fi{}kowski, Marek Trippenbach, and
  \L{}ukasz Cywi\ifmmode~\acute{n}\else\'{n}\fi{}ski.
\newblock Spectroscopy of cross correlations of environmental noises with two
  qubits.
\newblock {\em Phys. Rev. A}, 94:012109, Jul 2016.

\bibitem{Paz-Silva2017}
Gerardo~A. Paz-Silva, Leigh~M. Norris, and Lorenza Viola.
\newblock Multiqubit spectroscopy of gaussian quantum noise.
\newblock {\em Phys. Rev. A}, 95:022121, Feb 2017.

\bibitem{Krzywda2018}
Jan {Krzywda}, Piotr {Sza{\'n}kowski}, and {\L}ukasz {Cywi{\'n}ski}.
\newblock {The dynamical-decoupling-based spatiotemporal noise spectroscopy}.
\newblock {\em arXiv e-prints}, page arXiv:1809.02972, September 2018.

\bibitem{Slichter2010}
Charles~P. Slichter.
\newblock {\em Principles of magnetic resonance}.
\newblock Springer, 3rd edition, 2010.

\bibitem{Yoneda2018}
Jun Yoneda, Kenta Takeda, Tomohiro Otsuka, Takashi Nakajima, Matthieu~R.
  Delbecq, Giles Allison, Takumu Honda, Tetsuo Kodera, Shunri Oda, Yusuke
  Hoshi, Noritaka Usami, Kohei~M. Itoh, and Seigo Tarucha.
\newblock A quantum-dot spin qubit with coherence limited by charge noise and
  fidelity higher than 99.9\%.
\newblock {\em Nature Nanotechnology}, 13(2):102--106, 2018.

\bibitem{Petta2005}
J.~R. Petta, A.~C. Johnson, J.~M. Taylor, E.~A. Laird, A.~Yacoby, M.~D. Lukin,
  C.~M. Marcus, M.~P. Hanson, and A.~C. Gossard.
\newblock Coherent manipulation of coupled electron spins in semiconductor
  quantum dots.
\newblock {\em Science}, 309(5744):2180--2184, 2005.

\bibitem{Breuer2002}
H.~P. Breuer and F.~Petruccione.
\newblock {\em The theory of open quantum systems}.
\newblock Oxford University Press, Great Clarendon Street, 2002.

\end{thebibliography}

\end{document}